\documentclass[fleqn,twoside]{article}
\usepackage{espcrc2}

\usepackage{graphicx}
\usepackage{cite}
\usepackage{axodraw}
\usepackage{epsfig}

\newcommand{\lsim}
{\mathrel{\raisebox{-.3em}{$\stackrel{\displaystyle <}{\sim}$}}}

\def\asymp#1%
{\mathrel{\raisebox{-.4em}{$\widetilde{\scriptstyle #1}$}}}

\def\Nequal#1%
{\mathrel{\raisebox{-.5em}{$\stackrel{=}{\scriptstyle\rm#1}$}}}

\def\beq{\begin{equation}}
\def\eeq{\end{equation}}
\def\beqar{\begin{eqnarray}}
\def\eeqar{\end{eqnarray}}
\def\barr#1{\begin{array}{#1}}
\def\earr{\end{array}}
\def\bfi{\begin{figure}}
\def\efi{\end{figure}}
\def\btab{\begin{table}}
\def\etab{\end{table}}
\def\bce{\begin{center}}
\def\ece{\end{center}}

\def\text{\textstyle}

\def\Ga{\Gamma}
\def\ga{\gamma}

\def\si{\sigma}

\def\citere#1{\mbox{Ref.~\cite{#1}}}
\def\citeres#1{\mbox{Refs.~\cite{#1}}}

\newcommand{\TeV}{\unskip\,\mathrm{TeV}}
\newcommand{\GeV}{\unskip\,\mathrm{GeV}}
\newcommand{\MeV}{\unskip\,\mathrm{MeV}}

\newcommand{\fb}{\unskip\,\mathrm{fb}}

\newcommand{\rT}{{\mathrm{T}}}

\newcommand{\M}{{\cal{M}}}

\def\mathswitchr#1{\relax\ifmmode{\mathrm{#1}}\else$\mathrm{#1}$\fi}

\newcommand{\PW}{\mathswitchr W}

\newcommand{\Pg}{\mathswitchr g}

\newcommand{\PH}{\mathswitchr H}

\newcommand{\Pb}{\mathswitchr b}

\newcommand{\Pp}{\mathswitchr p}
\newcommand{\Pt}{\mathswitchr t}

\newcommand{\Pep}{\mathswitchr {e^+}}
\newcommand{\Pem}{\mathswitchr {e^-}}

\def\mathswitch#1{\relax\ifmmode#1\else$#1$\fi}

\newcommand{\Mt}{\mathswitch {m_\Pt}}
\newcommand{\Mg}{\mathswitch {m_\Pg}}

\def\solid{\raise.9mm\hbox{\protect\rule{1.1cm}{.2mm}}}
\def\dash{\raise.9mm\hbox{\protect\rule{2mm}{.2mm}}\hspace*{1mm}}

\def\ie{i.e.\ }

\title{NLO QCD corrections to
$\Pp\Pp{\to}\Pt\bar\Pt\Pb\bar\Pb{+}X$ 
via quark--antiquark annihilation%
\thanks{This work is supported by the European Community's Marie-Curie Research Training Network under contract MRTN-CT-2006-035505 ``Tools and Precision Calculations for Physics Discoveries at Colliders''.}
}

\author{A. Bredenstein\address[KEK]{High Energy Accelerator Research Organization (KEK),
Tsukuba, Ibaraki 305-0801, Japan}%
\thanks{Supported by the Japan Society for the Promotion of Science (JSPS).}, 
A. Denner\address[PSI]{Paul Scherrer Institut, W\"urenlingen und Villigen,
CH-5232 Villigen PSI, Switzerland}, 
S. Dittmaier\address[MPI]{Max-Planck-Institut f\"ur Physik
(Werner-Heisenberg-Institut),
D-80805 M\"unchen, Germany}
and S. Pozzorini\addressmark[MPI]}

\begin{document}

\begin{abstract}
The process $\Pp\Pp\to\Pt\bar\Pt\Pb\bar\Pb+X$ represents a very important
background reaction to searches at the LHC, in particular to 
$\Pt\bar\Pt\PH$ production where the Higgs boson decays into a
$\Pb\bar\Pb$ pair. A successful analysis of $\Pt\bar\Pt\PH$ at the LHC
requires the knowledge of direct $\Pt\bar\Pt\Pb\bar\Pb$ production
at NLO in QCD. We take the first step in this direction
upon calculating the NLO QCD corrections to the
subprocess initiated by $q\bar q$ annihilation. 
\vspace{1pc}
\end{abstract}

\maketitle

\section{Introduction}

The search for new particles will be the primary task of the LHC. 
The discovery of new particles
first of all requires to establish excess of events over
background. The situation at the LHC is particularly complicated by
the fact that for many expected signals the corresponding background
cannot entirely be determined from data, but has to be assessed upon
combining measurements in signal-free regions with theory-driven
extrapolations. To this end, a precise prediction for the background
is necessary, in particular including NLO 
corrections in QCD. Since many of these background processes involve
three, four, or even more particles in the final state, this kind of
background control requires NLO calculations at the technical
frontier. This problem lead to the creation of an ``experimenters'
wishlist for NLO calculations'' at the Les Houches workshop 2005
\cite{Buttar:2006zd}, updated in 2007 \cite{Bern:2008ef}, which
triggered great theoretical progress in recent years
(see \citere{Bern:2008ef} and references therein).
Meanwhile the listed processes involving at most five particles in loops
have been completed in NLO QCD, including the production of
$\PW\PW+\mathrm{jet}$~\cite{Dittmaier:2007th}, weak-boson pairs plus
two jets via vector-boson fusion~\cite{Jager:2006zc}, and triple
weak-boson production~\cite{Lazopoulos:2007ix,Hankele:2007sb}.
However, none of the true $2\to4$ processes has yet been addressed at
NLO.%
\footnote{
Progress in the calculation of the virtual corrections to
$\mathrm{u}\bar{\mathrm{u}}\to\mathrm{s}\bar{\mathrm{s}}\mathrm{b}\bar{\mathrm{b}}$
was reported in \citere{Binoth:2008gx}.}
Among those processes, $\Pp\Pp\to\Pt\bar\Pt\Pb\bar\Pb+X$ has top
priority.  

This  process  represents a very
important background to $\Pt\bar\Pt\PH$ production where the Higgs
boson decays into a $\Pb\bar\Pb$ pair.  While early studies of
$\Pt\bar\Pt\PH$ production at ATLAS~\cite{atlas-cms-tdrs} and
CMS~\cite{Drollinger:2001ym} suggested even discovery potential of
this process for a light Higgs boson, more recent
analyses~\cite{Cammin:2003,Cucciarelli:2006} with more realistic
background assessments show that the signal significance is
jeopardized if the background from $\Pt\bar\Pt\Pb\bar\Pb$ and
$\Pt\bar\Pt+\mathrm{jets}$ final states is not controlled very well.
This is a clear call for improved signal and background studies based
on NLO predictions to these complicated processes. For the
$\Pt\bar\Pt\PH$ signal~\cite{Beenakker:2001rj} and the
$\Pt\bar\Pt+\mathrm{1jet}$ background~\cite{Dittmaier:2007wz} at the
LHC such predictions have been accomplished in recent years.

The dominant mechanism to produce $\Pt\bar\Pt\Pb\bar\Pb$ final states
in hadronic collisions is pure QCD. In leading order (LO)
quark--antiquark ($q\bar q)$ and gluon--gluon ($\Pg\Pg$) initial
states contribute, where the latter strongly dominate at the LHC
because of the high gluon flux.  Being of order
$\alpha^4_{\mathrm{s}}$ the corresponding cross sections are affected
by a very large scale uncertainty, which amounts to a factor two or
more.  Technically the $q\bar q$ channel is simpler to deal
with---though still demanding---and thus represents a natural first
step towards a full treatment of $\Pp\Pp\to\Pt\bar\Pt\Pb\bar\Pb+X$ at
NLO. 
In these proceedings
we briefly report on this first step,
which was accomplished in  \citere{qqttbb}.

\begin{figure}
{\setlength{\unitlength}{0.50pt}
\SetScale{0.50}
\begin{picture}(160,90) (-80,-20)
\ArrowLine(-60,-35)(-20,-35)
\Vertex(-20,-35){2}
\ArrowLine(-20,-35)(-20,35)
\Vertex(-20,35){2}
\ArrowLine(-20,35)(-60,35)
\Gluon(-20,35)(30,35){4}{3.5}
\Gluon(-20,-35)(30,-35){4}{3.5}
\ArrowLine(60,15)(30,35)
\Vertex(30,35){2}
\ArrowLine(30,35)(60,55)
\ArrowLine(60,-15)(30,-35)
\Vertex(30,-35){2}
\ArrowLine(30,-35)(60,-55)
\end{picture}
\hspace*{1em}
\begin{picture}(160,90) (-90,-20)
\ArrowLine(-60,-40)(-40,0)
\Vertex(-40,0){2}
\ArrowLine(-40,0)(-60,40)
\Vertex(10,0){2}
\Gluon(-40,0)(10,0){4}{3.5}
\Gluon(10,0)(40,30){4}{3.5}
\Gluon(10,0)(40,-30){4}{3.5}
\ArrowLine(70,30)(40,30)
\Vertex(40,30){2}
\ArrowLine(40,30)(40,60)
\ArrowLine(70,-30)(40,-30)
\Vertex(40,-30){2}
\ArrowLine(40,-30)(40,-60)
\end{picture}
}
\caption{Sample diagrams
contributing to
$q\bar q\to\Pt\bar\Pt\Pb\bar\Pb$ in LO QCD.
}
\label{fig:LOtops}
\end{figure}
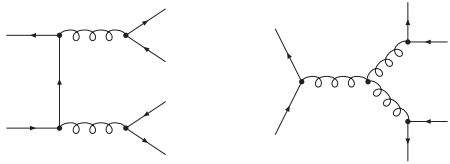
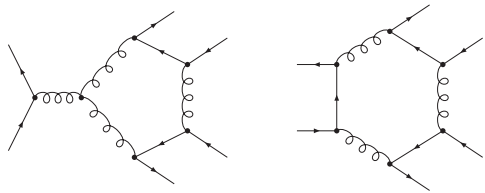
\begin{figure}
{\setlength{\unitlength}{0.50pt}
\SetScale{0.50}
\hspace*{1em}
\begin{picture}(170,70)(-90,-35)
\ArrowLine(-95,-40)(-75,0)
\Vertex(-75,0){2}
\ArrowLine(-75,0)(-95,40)
\Gluon(-75,0)(-40,0){4}{3.5}
\Vertex(-40,0){2}
\Gluon(-40,0)(0,45){4}{3.5}
\Gluon(-40,0)(0,-45){4}{3.5}
\Gluon(40,-25)(40,25){4}{3.5}
\ArrowLine(70,45)(40,25)
\Vertex(40,25){2}
\ArrowLine(40,25)(0,45)
\Vertex(0,45){2}
\ArrowLine(0,45)(30,65)
\ArrowLine(70,-45)(40,-25)
\Vertex(40,-25){2}
\ArrowLine(40,-25)(0,-45)
\Vertex(0,-45){2}
\ArrowLine(0,-45)(30,-65)
\end{picture}
\hspace*{1em}
\begin{picture}(160,70) (-80,-35)
\ArrowLine(-40,-25)(-40,25)
\Vertex(-40,25){2}
\ArrowLine(-40,25)(-70,25)
\Vertex(-40,-25){2}
\ArrowLine(-70,-25)(-40,-25)
\Gluon(-40,25)(0,50){4}{3.5}
\Gluon(-40,-25)(0,-50){4}{3.5}
\Gluon(40,-25)(40,25){4}{3.5}
\ArrowLine(70,45)(40,25)
\Vertex(40,25){2}
\ArrowLine(40,25)(0,50)
\Vertex(0,50){2}
\ArrowLine(0,50)(30,70)
\ArrowLine(70,-45)(40,-25)
\Vertex(40,-25){2}
\ArrowLine(40,-25)(0,-50)
\Vertex(0,-50){2}
\ArrowLine(0,-50)(30,-70)
\end{picture}}
\caption{Pentagon and hexagon sample graphs
contributing to $q\bar q\to\Pt\bar\Pt\Pb\bar\Pb$ at one loop in QCD.
}
\label{fig:NLOtops}
\end{figure}

In LO QCD seven different Feynman diagrams (Fig.~\ref{fig:LOtops})
contribute to the  partonic process
$q\bar q\to\Pt\bar\Pt\Pb\bar\Pb$. 
The virtual QCD corrections comprise about 200 one-loop diagrams, 
which include  8~hexagons and 24~pentagons
(Fig.~\ref{fig:NLOtops}).
The real corrections, $q\bar q\to\Pt\bar\Pt\Pb\bar\Pb \Pg$, 
are described by  64
diagrams, which result
from the LO graphs upon adding an external 
gluon in 
all possible ways.
In the following we briefly describe the calculation of the virtual
and real corrections and present numerical results for the LHC.
The calculation has been 
worked out twice and independently, resulting in two completely
independent computer codes, which we denote as version 1 and 2.

\section{Virtual corrections}

The general strategy for the evaluation of the one-loop corrections
is based on the reduction of the amplitude $\M^{(\Ga)}$ of each 
(sub)diagram $\Ga$  in the following way,
\beq\label{colspinreduction}
\M^{(\Ga)} = 
 {\cal C}^{(\Ga)}
\left(\sum_m {\cal F}^{(\Ga)}_m
\, \hat\M_m
\right),
\eeq
where the colour structure ${\cal C}^{(\Ga)}$ 
present in the (sub)diagram
is factorized 
from the remaining colour-independent part.
While loop diagrams that involve a quartic gluon vertex
give rise to $3$ colour-factorized
amplitudes $\M^{(\Ga)}$, in all other diagrams---which constitute the large majority---the colour structure factorizes completely.
This implies that the computation time for individual loop diagrams does
not scale with the total number of independent colour structures, 
which is 6 for $q\bar q\to\Pt\bar\Pt\Pb\bar\Pb$.

The colour-free part of $\M^{(\Ga)}$ 
is  written as a linear combination of so-called standard matrix elements
(SMEs)  $\hat\M_m$, 
which consist of Dirac chains containing 
the polarization information. 
Since the computing time scales with the number of 
SMEs, it is important to reduce the set of SMEs 
as much as possible.
To this end, we employ an algebraic procedure 
based on four-dimensional relations that are 
derived from Chisholm's identity 
and are applied after separation of
UV and IR singularities.
For massless external
fermions this four-dimensional reduction has been described in detail
in Sections~3.1 and 3.3 of \citere{Denner:2005fg}; here we had to
generalize this approach to one massive and two massless spinor
chains (b quarks are treated as massless particles).
In this way some thousand different spinor chains are reduced
to about 150 SMEs 
$\hat\M_m$.

The most time-consuming components of the
numerical calculation are the
scalar form factors ${\cal F}^{(\Ga)}_m$,
which are linear combinations of the 
Lorentz-invariant coefficients of  
$N$-point  tensor loop integrals
with rank $R \le 3$ and degree $N\le 6$,
\beqar
{\cal F}^{(\Ga)}_m
&=& 
\sum_{R} \,
\sum_{j_1,\dots,j_R}\,
{\cal K}^{(\Ga)}_{m;j_1,\dots,j_R}
T^{N}_{j_1,\dots,j_R}
.
\eeqar
The evaluation of the tensor coefficients $T^{N}_{j_1,\dots,j_R}$
follows the strategy of \citeres{Denner:2002ii,Denner:2005nn}
that was already successfully used to compute the NLO electroweak
corrections to $\Pep\Pem\to4\,$fermions \cite{Denner:2005fg,Denner:2005es}.
In this approach 
the analytic expressions are not reduced to master integrals.
In contrast, the tensor integrals are evaluated by means of algorithms that 
perform a recursive reduction to master integrals in numerical form.
This avoids huge analytic expressions and
permits to adapt the reduction strategy
to the specific numerical problems that 
appear in different phase-space regions.
The scalar master integrals are evaluated 
using the methods and results of \citeres{'tHooft:1979xw,Beenakker:1990jr}. 
UV divergences are regularized dimensionally in both evaluations, but
IR divergences are treated in different ways as described below.
Tensor and scalar 6-/5-point
functions are directly expressed in terms of 5-/4-point integrals
\cite{Denner:2002ii,Denner:2005nn}.
Tensor 4-point and 3-point
integrals are reduced to scalar integrals with the Passarino--Veltman
algorithm \cite{Passarino:1979jh} as long as no small Gram determinant
appears in the reduction. 
If small Gram determinants occur, two alternative
schemes are applied \cite{Denner:2005nn}.
One method makes use of
expansions of the tensor coefficients about the limit of vanishing
Gram determinants and possibly other kinematical determinants.  In the
second (alternative) method we evaluate a specific tensor coefficient,
the integrand of which is logarithmic in Feynman parametrization, by
numerical integration. Then the remaining coefficients as well as the
standard scalar integral are algebraically derived from this
coefficient.

{\it Version~1} of the virtual corrections starts with the generation
of Feynman diagrams using {\sc FeynArts}~1.0~\cite{Kublbeck:1990xc}.
Their algebraic reduction is completely performed with in-house {\sc
  Mathematica} routines. In detail, $D$-dimensional identities (Dirac
algebra, Dirac equation) are used until UV divergences cancel against
counterterms. 
The  IR (soft and collinear) divergences
are regularized
dimensionally and 
expressed in terms
of scalar 2- and 3-point
integrals 
as described in \citere{Dittmaier:2003bc}, keeping the 
full dependence on $D$.
It turns out that no $D$-dependent prefactors occur in front of
the IR-singular $B_0$ and $C_0$ integrals. In \citere{qqttbb}
we show that this
result of our specific calculation is not accidental, but generalizes
to arbitrary processes at NLO in the Feynman gauge.
Having cancelled UV divergences against counterterms and
controlled the $D$-dimensional issues concerning IR singularities, the
amplitude is further simplified by reducing the SMEs
in four space--time dimensions.

{\it Version~2} of the virtual corrections employs {\sc
  FeynArts}~3.2~\cite{Hahn:2000kx} for generating and {\sc
  FormCalc}~5.2~\cite{Hahn:1998yk} for preprocessing the amplitudes.
The first part of the calculation is performed in $D$ dimensions.  In
particular, the so-called rational terms resulting from the UV
divergences of tensor loop coefficients are automatically extracted by
{\sc FormCalc}.  Since the IR divergences that appear in the $q\bar q$
channel are of abelian nature, we exploit the fact that they can be
regularized as in QED by means of fermion and gauge-boson (gluon)
masses, $m_q$ and $\Mg$.  These masses are treated as infinitesimal
quantities (with $\Mg\ll m_q$) both in the algebraic expressions and
in the numerical routines that evaluate the tensor integrals, \ie only
the logarithmic dependence on these mass parameters is retained.
Corresponding IR singularities associated with real emission have been
obtained from \citere{Catani:2002hc} by means of an appropriate change
of regularization scheme.

\section{Real corrections}

In both evaluations 
the singularities from soft or
collinear gluon emission are isolated via dipole
subtraction~\cite{Catani:2002hc,Catani:1996vz,Dittmaier:1999mb} 
for NLO QCD
calculations using the formulation~\cite{Catani:2002hc} for massive quarks.
After combining virtual and real corrections, singularities connected to
collinear configurations in the final state cancel for collinear-safe
observables automatically after applying a jet algorithm. Singularities
connected to collinear initial-state splittings are removed via
factorization by PDF redefinitions.  
The phase-space integration is performed with multichannel Monte Carlo
generators \cite{Berends:1994pv} and adaptive weight optimization
similar to the one implemented in {\sc RacoonWW} \cite{Denner:1999gp}.

In {\it version~1} of the real corrections the matrix elements have
been calculated  using the 
Weyl--van-der-Waerden spinor technique in the formulation of
\citere{Dittmaier:1998nn}. Soft and collinear singularities are
regularized using dimensional regularization. 
The phase-space integration, implemented in {\sc C++}, is based
on {\sc RacoonWW}, but the phase-space mappings are build up in a more generic
way very similar to the approach of {\sc Lusifer}~\cite{Dittmaier:2002ap}.

In {\it version~2} of the real corrections the matrix elements have
been generated with {\sc Madgraph} \cite{Stelzer:1994ta}. As in the
virtual corrections, soft singularities are regularized by an
infinitesimal gluon mass and collinear singularities by small quark
masses, which appear only in logarithms in the endpoint part of the 
subtraction function.
The Monte Carlo generator is a further development of the one used in
{\sc COFFER$\ga\ga$} \cite{Bredenstein:2005zk} and for the calculation
of the NLO corrections to $\Pp\Pp\to\mathrm{jj}\PH+X$
\cite{Ciccolini:2007jr}.
For the purpose of checking, we have also implemented two-cut-off slicing. 
In this approach the singular parts of the 
phase-space integration, which originate from the soft region
$ E_\Pg < \delta_\mathrm{s} \sqrt{\hat s}/2$
and the collinear regions
$1-\cos(\theta_{\Pg q}) < \delta_\mathrm{c}$,
are treated by means of the 
well-known 
soft/collinear approximations.
When adding all contributions, the
${\cal O}(\delta_\mathrm{s,c})$ 
dependence on the technical cuts cancels if the cut-off parameters are
chosen small enough.

\section{Numerical results}
\label{se:numres}
\begin{figure}
\includegraphics
[bb= 85 440 285 660, width=.35\textwidth ]
{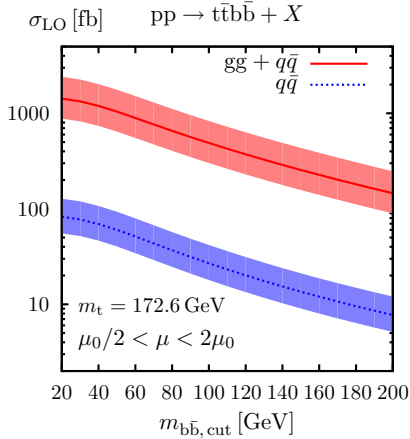}
\vspace{-8mm}
\caption{Complete LO cross section ($\Pg\Pg+q\bar q$)
and $q\bar q$ contribution as a function
of $m_{\Pb\bar\Pb,\mathrm{cut}}$.
}
\label{fig:LOcsmbb}
\end{figure}
\begin{figure}
\includegraphics[bb= 85 440 285 660, width=.35\textwidth]{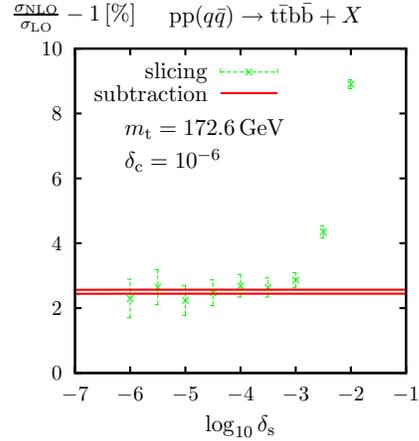}
\vspace{-8mm}
\caption{NLO corrections 
to $\Pp\Pp(q\bar q)\to\Pt\bar\Pt\Pb\bar\Pb+X$:
dipole subtraction versus phase-space slicing.}
\label{fig:subsli}
\end{figure}

We present predictions for
$\Pp\Pp\to\Pt\bar\Pt\Pb\bar\Pb+X$ at $\sqrt{s}=14\TeV$. 
For the top-quark mass
we take $\Mt=172.6\GeV$, while 
all other QCD partons (including $\Pb$~quarks)
are treated as massless.
Final states involving 
collinear gluons and b-quarks are recombined into 
collinear-safe jets by means of
the \mbox{$k_\rT$-algorithm} of \citere{Blazey:2000qt}.
Specifically we require two b-quark
jets with separation $\sqrt{\Delta\phi^2+\Delta
  y^2}>D=0.8$ in the rapidity--azimuthal-angle plane.
This sets an effective lower limit on the
invariant mass $m_{\Pb\bar\Pb}$ of the b-quark pair,
thereby avoiding collinear singularities from 
$\Pg\to\Pb\bar\Pb$ splittings.
Motivated by the
search for a $\Pt\bar \Pt H (H\to\Pb\bar \Pb)$ signal at the LHC
\cite{Cammin:2003,Cucciarelli:2006}, we impose the following additional
cuts on the transverse momenta, rapidity and 
invariant mass of the b-quark jets:
\mbox{$p_{\rT,\Pb}>20\GeV$}, \mbox{$|y_\Pb|<2.5$}
and $m_{\Pb\bar\Pb}>m_{\Pb\bar\Pb,\mathrm{cut}}$.
We use CTEQ6  PDFs~\cite{Pumplin:2002vw}
with  $\Lambda_5^{\mathrm{LO}}=165\MeV$ at LO 
and $\Lambda_5^{\overline{\mathrm{MS}}}=226\MeV$ at NLO.
For the renormalization and factorization scales 
we use the central value $\mu_0=\Mt+m_{\Pb\bar\Pb,\mathrm{cut}}/2$
and, to estimate the scale uncertainty, we vary
these scales in the range $\mu_0/2<\mu<2\mu_0$ 
in a uniform and antipodal way, \ie
with $\mu_{\mathrm{R}}=\mu_{\mathrm{F}}=\mu$
and
$\mu_{\mathrm{R}}=\mu_0^2/\mu_{\mathrm{F}}=\mu$.

\begin{figure}
\includegraphics[bb= 85 440 285 660, width=.35\textwidth]{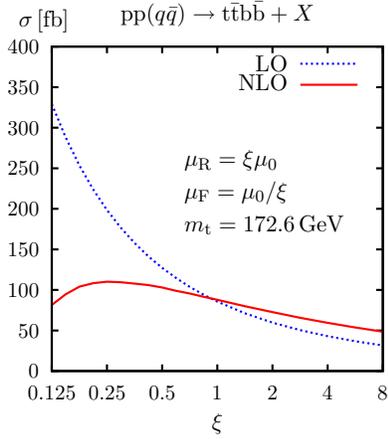}
\vspace{-8mm}
\caption{ LO and NLO scale dependence  of 
$\Pp\Pp(q\bar q)\to\Pt\bar\Pt\Pb\bar\Pb+X$ 
(antipodal scale variation).
}
\label{fig:scaledep}
\end{figure}

\begin{figure}
\includegraphics[bb= 85 440 285 660, width=.35\textwidth]{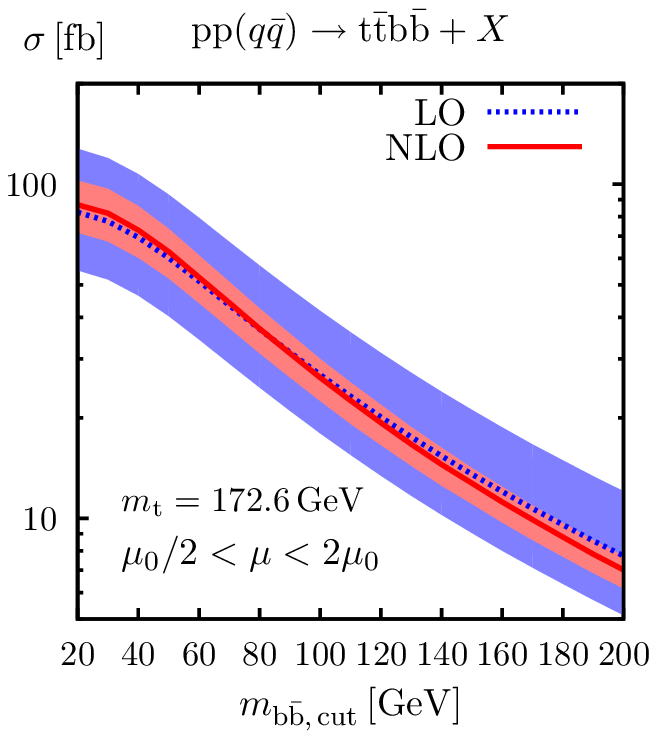}
\vspace{-8mm}
\caption{LO and NLO 
$\Pp\Pp(q\bar q)\to\Pt\bar\Pt\Pb\bar\Pb+X$ 
cross section 
as a function of $m_{\Pb\bar\Pb,\mathrm{cut}}$.}
\label{fig:NLOcsmbb}
\end{figure}

Figure~\ref{fig:LOcsmbb} shows the total LO cross section and the
contribution induced by $q\bar q$ annihilation as a
function of 
$m_{\Pb\bar\Pb,\mathrm{cut}}$.
The $\Pg\Pg$ channel dominates over $q\bar q$ annihilation
by roughly a factor of 17.  
The scale uncertainty,
displayed by the bands,
amounts to a factor 1.6 and
is dominated by the $\mu_{\mathrm{R}}$-dependence
owing to the large power of
$\alpha_{\mathrm{s}}(\mu_{\mathrm{R}})^4$ in the LO cross section.
Figure~\ref{fig:subsli} illustrates the mutual agreement between 
NLO results for the  $q\bar q$ channel
obtained with dipole subtraction and phase-space slicing.
Here we set $m_{\Pb\bar\Pb,\mathrm{cut}}=0$, \ie $\mu_0=\Mt$.
We observe that, within integration errors, the slicing results become
independent of the soft cut-off $\delta_\mathrm{s}$
for 
$\delta_\mathrm{s}\lsim 10^{-3}$ and
agree with the result in
the subtraction method. 
For the  $\delta_\mathrm{c}$-dependence we
find a similar behaviour~\cite{qqttbb}.
Figure~\ref{fig:scaledep} shows
the reduction of the
scale dependence of the $q\bar q$ contribution
upon going
from LO to NLO. 
Setting $m_{\Pb\bar\Pb,\mathrm{cut}}=0$,
at  $\mu=\mu_0=\Mt$ we find 
$\si_{\mathrm{LO}}=85.522(26)\fb$ and
$\si_{\mathrm{NLO}}=87.698(56)\fb$,
and we observe that (uniform and antipodal) 
scale variations in the range
$\mu_0/2<\mu<2\mu_0$ shift
the cross section by 55\% in LO and by 17\% in NLO. 
Figure~\ref{fig:NLOcsmbb} displays the dependence 
of the LO and NLO cross sections on 
the invariant-mass cut $m_{\Pb\bar\Pb,\mathrm{cut}}$.
The central scale and the uncertainty 
bands are as in Figure~\ref{fig:LOcsmbb}.
The reduction of the scale uncertainty from about $\pm50\%$ to
$\pm17\%$ holds true for the
considered range in $m_{\Pb\bar\Pb,\mathrm{cut}}$.
While the NLO prediction is consistent with the LO uncertainty band,
the shape of the distribution is distorted by the corrections. 
For the
central scale,  $\mu_0=\Mt+m_{\Pb\bar\Pb,\mathrm{cut}}/2$,
we find an NLO correction of $+2.5\%$ for small
$m_{\Pb\bar\Pb,\mathrm{cut}}$ but a correction of $-10.6\%$ for
$m_{\Pb\bar\Pb,\mathrm{cut}}=200\GeV$.

\section{Numercial stability and CPU efficiency}
In order to validate the numerical stability of 
the tensor reduction we compared the results
of the two  independent codes finding
very good agreement both for
single phase-space points and
integrated quantities.
The CPU time needed to evaluate the virtual corrections for a
phase-space point (including sums over colours and polarizations)
amounts to about 10\,ms on a 3\,GHz Intel Xeon processor.
This demonstrate that the employed diagrammatic techniques 
permit a fast and numerically stable
evaluation  of six-particle processes
at the LHC.
Based on these encouraging results, we expect to be able to extend
this calculation to the technically more challenging gluon-fusion
channel.

\end{document}